\documentclass[fleqn,usenatbib]{mnras}
\usepackage{newtxtext,newtxmath}
\pdfoutput=1 
\usepackage[T1]{fontenc}
\usepackage[utf8]{inputenc}
\usepackage{ae,aecompl}
\usepackage{float}

\usepackage{graphicx}	
\usepackage{amsmath}	
\usepackage{amssymb}	
\usepackage[usenames]{xcolor}
\usepackage{color}
\usepackage{booktabs}
\usepackage{footnote}
\makesavenoteenv{tabular}
\usepackage[flushleft]{threeparttable}

\title[A proxy for the gas disc mass]{Multi-wavelength observations of protoplanetary discs as a proxy for the gas disc mass}

\author[Veronesi et al.]{
B. Veronesi,$^{1,3}$\thanks{E-mail: benedetta.veronesi@unimi.it}
G. Lodato,$^{1,3}$
G. Dipierro,$^{2}$
E. Ragusa,$^{2}$
C. Hall$^{2,3}$
and D.~J. Price$^{3}$
\\
\\
$^{1}$Dipartimento di Fisica, Universit\`a degli Studi di Milano, Via Celoria, 16, Milano, I-20133, Italy\\
$^{2}$Department of Physics and Astronomy, University of Leicester, Leicester, United Kingdom\\
$^{3}$School of Physics and Astronomy, Monash University, VIC 3800, Australia
}

\date{Accepted XXX. Received YYY; in original form ZZZ}

\pubyear{2019}

\begin{document}
\label{firstpage}
\pagerange{\pageref{firstpage}--\pageref{lastpage}}
\maketitle

\begin{abstract}
Recent observations of protoplanetary discs reveal disc substructures potentially caused by embedded planets. We investigate how the gas surface density in discs changes the observed morphology in scattered light and dust continuum emission. Assuming that disc substructures are due to embedded protoplanets, we combine hydrodynamical modelling with radiative transfer simulations of dusty protoplanetary discs hosting planets. The response of different dust species to the gravitational perturbation induced by a planet depends on the drag stopping time --- a function of the generally unknown local gas density. Small dust grains, being stuck to the gas, show spirals. Larger grains decouple, showing progressively more axisymmetric (ring-like) substructure as decoupling increases with grain size or with the inverse of the gas disc mass. We show that simultaneous modelling of scattered light and dust continuum emission is able to constrain the Stokes number, ${\rm St}$. Hence, if the dust properties are known, this constrains the local gas surface density, $\Sigma_{\rm gas}$, at the location of the structure, and hence the total gas mass. In particular, we found that observing ring-like structures in mm-emitting grains requires ${\rm St} \gtrsim 0.4$ and therefore $\Sigma_{\rm gas} \lesssim 0.4\,\textrm{g/cm}^{2}$. We apply this idea to observed protoplanetary discs showing substructures both in scattered light and in the dust continuum.
\end{abstract}




\section{Introduction}

The exceptional imaging capabilities of new instruments such as the Spectro-Polarimetric High-contrast Exoplanet REsearch (SPHERE) imager at the Very Large Telescope and observatories like the Atacama Large Millimeter Array (ALMA) offer exciting new possibilities to explore the dynamics of the two main components of protoplanetary discs: gas and dust.
High-resolution observations at a wide range of wavelengths have revealed disc substructures.
These substructures are common, as a result of ubiquitous processes in disc evolution \citep{garufi18a}. Importantly, the morphology of these substructures may differ in scattered light ($\mu$m) observations compared to dust thermal emission at $\sim$mm wavelengths \citep[e.g.][]{follette13a,casassus16a,van-boekel16a,hendler17a,pinilla18a,dong18a}.
While scattered light observations at near-infrared (NIR) wavelengths trace dust particles with sizes of up to a few microns in the disc surface layers (i.e. the scattering surface), continuum emission at (sub)-millimetre wavelengths provides one of the most direct constraints on the spatial distribution of millimeter-sized dust grains in the discs midplane. Therefore, the different morphology of disc substructures at different wavelengths may be due to the different dynamics of small and large dust grains, making the combination of scattered light (with, e.g., SPHERE) and continuum thermal observations (with, e.g. ALMA) a powerful tool to explore the gas and dust dynamics in protoplanetary discs.

For example, the structures of the disc around HD135344B are different in scattered light observations \citep{maire17} with respect to the dust continuum \citep{cazzoletti18}. While in the first case we detect a two armed spiral structure, in the second one we observe an inner ring and an asymmetric structure. It is also interesting to note that by looking at the dust continuum at different wavelengths this asymmetry can be described as part of a ring (in Band 3 and Band 4, see \citealt{cazzoletti18}) or as an horseshoe (in Band 9 and Band 7). Assuming that these structures are generated by a planet (though it has been argued that they could have been originated by other mechanisms, like vortices, e.g. \citealt{vandermarel16}), we know that gas and dust respond differently to the presence of a massive, Jupiter-like planet, with a distinctive spiral structure in the gas (and the small dust tightly coupled to it) \citep[e.g.][]{dong15b}, but a more axisymmetric ring-like structure in large dust grains  \citep[e.g.][]{ayliffe12a,dong15a,dipierro15b,price18}. 

The morphology of structures observed at different wavelengths depends on the interaction between the dust and gas. 
The motion of dust particles is affected by the interaction with the gas with an efficiency that depends on the degree of aerodynamical coupling. 
This is quantified by the Stokes number, St, the ratio of the drag stopping time $t_{\rm s}$ and the dynamical time in the disc, i.e. \citep[e.g.][]{weiden77}
\begin{equation}
\mathrm{St} \equiv t_{\rm s}\Omega_{\mathrm{k}} =\sqrt{\frac{\pi}{8}} \frac{a \rho_{\rm d} \Omega_{\mathrm{k}}}{\rho_{\rm{gas}} c_{\mathrm{s}}}=\frac{\pi}{2} \frac{a \rho_{\rm d}}{\Sigma_{\rm{gas}}}\exp{\left(\frac{z^2}{2H^2_{\rm{g}}}\right)},
\label{eq:stokes}
\end{equation}
where $\Omega_{\mathrm{k}}$ is the Keplerian angular velocity, $\rho_{\rm d}$ is the internal dust density, $z$ is the vertical coordinate, $a$ is the size of dust grains, $\Sigma_{\rm{gas}}$ is the gas surface density, $H_{\rm{g}}$ is the disc scale height assuming vertical hydrostatic equilibrium, i.e. $H_{\rm{g}}=c_{\mathrm{s}}/\Omega_{\mathrm{k}}$ where $c_{\mathrm{s}}$ is the sound speed. The above equation is valid for small dust-to-gas mass ratio, isothermal equation of state and in the Epstein drag regime \citep{epstein24a}, which is the relevant regime of dust-gas coupling up to millimeter-centimetre dust grains for typical disc parameters \citep{garaud04a}. Strongly coupled and weakly coupled dust is defined by having $\mathrm{St}\ll 1$ and $\mathrm{St}\gg 1$, respectively. Moreover, as we can see from Eq.~\ref{eq:stokes}, the coupling between dust and gas decreases with increasing vertical distance from the midplane. For this reason, 3D simulations of dust and gas are necessary in order to study the dynamics of the disc and to understand if this different degree of coupling can be detected by SPHERE (tracing grains at $z\approx 2-3$ scale heights) and by ALMA (tracing grains at $z \sim 0$). The Stokes number depends on an elusive yet fundamental disc property --- the gas surface density $\Sigma_{\rm{gas}}$, and thus the total gas disc mass.

The total gas disc mass, $M_{\rm gas}$, determines how much mass is available for the formation of giant planets and for the accretion of gas atmosphere around rocky planets. At the same time, it also determines the likelihood of gravitational instability, which may induce disc fragmentation and enhanced angular momentum transport \citep{kratter16}. Correlations of the disc mass with either the stellar mass \citep{natta2000,andrews13,andrews18a} or the mass accretion rate \citep{mulders17a,lodato17} can provide important constraints on disc evolution. 

The dust mass is relatively easy to measure, starting from optically thin observations at millimetre wavelengths, provided that we have adequate information on the dust opacity and thus on the level of dust growth \citep{bergin18a}. However, also in this case there is a lot of uncertainty, due to the optical depth estimate of the dust at (sub-)mm wavelengths. \cite{woitke16} discuss this in terms of the uncertainties we have in the dust grain size and composition (see their Fig.3, respectively roughly half and one order of magnitude).
For decades, the total disc mass has been obtained from the dust mass simply by assuming a gas-to-dust ratio of 100, typical of the interstellar medium \citep{mathis77}. Whether this is applicable to the dense environments of protoplanetary discs is not yet understood \citep[e.g.][]{bate17a}. Moreover, recent observations of protoplanetary discs have found a discrepancy between dust and gas disc sizes \citep[e.g.][]{ansdell16a,ansdell18a}, that can be explained by the dynamical effect of the dust radial drift \citep{birnstiel14a}. The different dynamics between the two phases inevitably create regions of enhanced or reduced dust-to-gas ratio with respect to its ISM value.

Direct gas tracers are more difficult. The main component in the gas is the H$_2$ molecule, which lacks a permanent electric dipole and so does not emit significantly. 
A proxy for the gas mass is provided by observations of CO in its various isotopologues (such as $^{13}$CO and  C$^{18}$O), although it is not clear whether these measurements are a reliable estimate of the gas disc mass \citep{williams14a,bergin18a}. This is due to the fact that the conversion of the observed CO mass into total gas mass is not well understood. For example, the above mentioned correlations between disc mass and either stellar mass or accretion rates are much stronger when one uses dust rather than CO as a proxy for the disc mass \citep{manara16a}. Typically, CO observations result in very low disc masses compared to dust estimates \citep{pascucci16a,ansdell16a,miotello17,long17a}, which might indicate substantial carbon depletion in the disc, photodissociation in the upper layers, freeze-out at the disc midplane  or in general other isotope-selective processes \citep{miotello16a}. Measurements of the HD line in the far infrared \citep{bergin13} have proved even more controversial, suggesting very high gas masses \citep[but see][]{trapman17}.
Importantly, \cite{manara2018} found that not only for the gas mass estimate, but also for the dust mass the picture is not trivial. Indeed, they found that measurements of dust mass of protoplanetary discs in $\sim 1-3$ Myr old regions are lower than the core masses in exo-planets and planetary systems. 

In this paper, we propose a method to infer the gas surface density co-located with the continuum and scattered light emissions of protoplanetary discs showing evidence of planet-induced substructures. We base our analysis on the difference in morphology of disc substructures imaged in NIR scattered light and (sub-)mm continuum observations. By combining 3D numerical simulations of a suite of dusty protoplanetary disc models hosting embedded protoplanets with 3D Monte Carlo radiative transfer simulations, we analyse the different observational predictions of disc substructures imaged by SPHERE and ALMA. The main aim is to find an empirical method to link the different morphology of disc substructures at different wavelengths with the gas surface density. 
This paper is organised as follows: in Section~\ref{sec:numerics} we describe our numerical method and simulation setup. In Section~\ref{sec:results} we describe the results of the numerical simulations and show a set of synthetic SPHERE and ALMA images of disc model. Finally, in Section~\ref{sec:discussion} we discuss the implications of our results for the estimate of the gas surface density in protoplanetary discs and draw our conclusions in Section~\ref{sec:conclusion}.

\section{Methods}
\label{sec:numerics}

\subsection{Dust and gas numerical simulations}
We perform a suite of 3D Smoothed Particle Hydrodynamics (SPH) simulations of dusty protoplanetary discs, using the \textsc{phantom} code developed by \cite{price18phantom}. Depending on the level of aerodynamical coupling between the gas and dust, we adopt the one fluid (for St$<1$, \citealt{price15,ballabio18}) or the two fluid (St$>1$, \citealt{laibe12a,laibe12b}) methods to simulate the dynamics of dust grains.
The one-fluid algorithm is based on the terminal velocity approximation (e.g. \citealt{youdin05}). In all our simulations, we include the dust back-reaction,
but do not include self-gravity.

\begin{figure*}
\includegraphics[scale=0.6,trim={0 12.cm 0 0},clip]{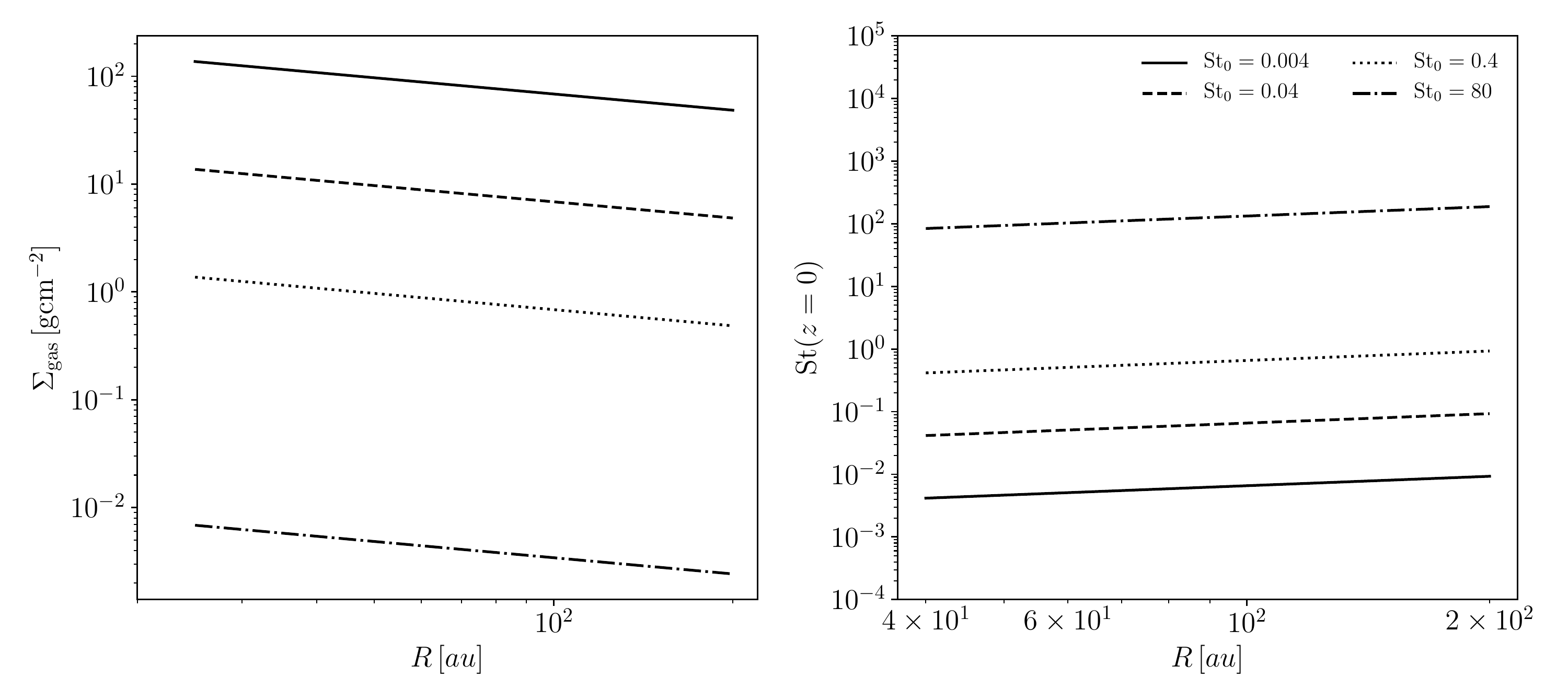}
    \caption{Gas surface density (left) and midplane Stokes number (right) in our simulations as a function of radius for different midplane Stokes number (at the inner dust radius) $\rm{St}_0 \approx 4\cdot[10^{-3}, 10^{-2}, 10^{-1}, 20]$}. 
    \label{fig:stokes}
\end{figure*}

\subsection{Disc models}
We pick as a test object HD135344B (also known as SAO 206462), a young stellar object hosting a dusty protoplanetary disc showing a prominent two-armed spiral structure in the near-IR scattered light emission \citep{stolker16,maire17} and an inner ring plus a large horseshoe feature at (sub-)mm wavelengths \citep{vandermarel16,cazzoletti18}. HD135344B is also known to be part of a visual binary, but due to the wide separation between the two objects ($21''$, i.e. $\simeq 3280$ au, \citealt{mason01}) it is unlikely that the disc structure we see is originated from the interaction with the companion. 
This motivates the choice of the parameters for the initial disc conditions in our SPH simulations (outlined in Table~\ref{tab:setups}), although we note that the aim of our study is more generally valid to different configurations of planet-disc interaction.  

In order to pursue this aim, in our hydrodynamical simulations, rather than changing directly the gas density, we change the Stokes number to obtain a corresponding variation of the gas mass: for example, since St is inversely proportional to the gas density (Eq. \ref{eq:stokes}), an increase in St by a factor of 10 is equivalent to a decrease in $\Sigma_{\rm gas}$ by the same factor. This is needed because we want to compare disc morphologies independently of other evolutionary effects that might be affected by the gas mass, such as planetary migration and accretion \citep[e.g.][]{kley12a,dangelo08a}. 
Once the Stokes number and the dust properties (i.e. fluffiness, porosity, composition) are known, it is always possible to go back to the gas mass that has generated that given gas and dust coupling and the resulting system dynamics. 

\begin{table}
\caption{Model parameters. St$_0$ is the midplane Stokes number at the dust inner radius $R_{\rm in,d}$, $R_{\rm in,g}$ is the gas inner radius. $p$ is the power-law index of the gas surface density profile defined in Eq.~\ref{eq:power}, $q$ is the power-law index of the sound speed radial profile (see Eq.~\ref{eq:soundspeed}) and $\alpha_{\rm ss}$  is the effective \citet{shakura73} viscosity. $M_{\rm IP,OP}$ and  $R_{\rm IP,OP}$ are the mass of the inner and outer planets and their radial position in the disc.}
\label{tab:setups} 
\begin{center}
\begin{tabular}{lc} 
 \hline
 \hline
Parameters &  Value \\ 
 \hline

$M_\star\, [M_\odot]$ & $1.7$ \\
$R_{\rm in,g}\, [{\rm au}]$  & $25$ \\
$R_{\rm in,d}\, [{\rm au}]$  & $40$ \\
$R_{\rm out,g}=R_{\rm out,d} \, [{\rm au}]$ & $200$ \\
$(H/R_{\rm in})_{\rm g}$ & $0.04,0.06,0.08$ \\
$M_{\rm dust}\, [M_\odot]$ & $5.2 \cdot 10^{-4}$ \\
$ \rm{St}_0 $ (approx) & $4\cdot[10^{-3}, 10^{-2}, 10^{-1}, 20]$ \\
$\rho_{\mathrm{d}}\, [{\rm g\,cm}^{-3}]$ &$3$ \\
$p$ & $0.5$\\ 
$q$ & $0.35$\\
$\alpha_{\rm SS}$  & $ 0.007$    \\
\hline
$M_{\rm IP}\,[M_{\mathrm{j}}]$   & $3$\\
$R_{\rm IP}\,[{\rm au}]$  & $ 35 $ \\
$M_{\rm OP}\,[M_{\mathrm{j}}]$   & $5$\\
$R_{\rm OP}\,[{\rm au}]$  & $ 145 $ \\ 
\hline
\end{tabular}
\end{center}
\end{table}

\subsubsection{Gas and dust}
\label{sec:gasdust_model}
The system consists of a central star of mass $M_\star=1.7\, M_{\odot}$ surrounded by a gas disc extending from $R_{\rm in,g}$ = 25 au to $R_{\rm out,g}$ = 200 au and modelled as a set of $10^{6}$ SPH particles. 
The initial gas surface density profiles are assumed to be power laws \citep{fung15}, i.e.
\begin{equation}
\Sigma_{\rm gas}(R)= \Sigma_{\rm in} \left(\frac{R}{R_{\rm in,g}}\right)^{-p}  ,
\label{eq:power}
\end{equation} 
where $\Sigma_{\rm in}$ is a normalization constant at the inner radius and $p=0.5$. We adopt a locally isothermal equation of state $P=c_{\rm s}^{2} \rho_{\mathrm{g}}$, with
\begin{equation}
c_{\mathrm{s}} = c_{\mathrm{s},{\rm in}} \left(\frac{R}{R_{\rm in,g}} \right)^{-q} ,
\label{eq:soundspeed}
\end{equation}
where $c_{\mathrm{s},{\rm in}}$ is the sound speed at the inner disc radius 
and $\rho_{\mathrm{g}}$ is the gas volume density. We assume $q=0.35$ \citep{andrews11,carmona14} as the power-law index of the sound speed radial profile. 
The disc is vertically extended by assuming a Gaussian profile for the volume density and ensuring vertical hydrostatic equilibrium 
\begin{equation}
\frac{H_{\rm g}}{R} = \frac{c_{\mathrm{s}}}{v_{\mathrm{k}}}=\left(\frac{H}{R_{\rm in}}\right)_{\rm g}\left(\frac{R}{R_{\rm in,g}}\right)^{1/2-q}  ,
\label{eq:aspectratio}
\end{equation}
where $v_{\mathrm{k}}$ is the Keplerian velocity and $(H_{\rm g}/R)_{\rm in}$ is the aspect ratio at the reference radius $R_{\rm in}$. 
In our study, we vary the aspect ratio $(H/R_{\rm in})_{\rm g}$ in the range $[0.04,0.06,0.08]$ \citep{stolker16}. 
We model viscous gas discs with an effective \citet{shakura73} viscosity $\alpha_{\mathrm{SS}} \approx 0.007$, implemented through the artificial viscosity formalism in SPH \citep{lodato10a}.

Given the different cavity radius in gas and dust observed in HD135344B disc \citep{garufi13,vandermarel15}, we consider a dust disc extending from $R_{\rm in,d}=40$ au to $R_{\rm out,d}=200$ au. 
We use the same functional form of the initial surface density as for the gas (Eq.~\ref{eq:power}), assuming a dust mass of $M_{\mathrm{dust}}=5.2 \cdot 10^{-4}\, M_\odot$. The dust-to-gas is initially assumed constant for the whole disc extent, so that the dust has the same vertical structure of the gas. After a few orbits of the outer planet, the dust has settled down forming a layer with thickness $H_{\rm d}=H_{\rm g}\sqrt{\alpha_{\rm ss}/({\rm St}+\alpha_{\rm ss})}$ \citep{fromang09}.

\subsubsection{Treating different degrees of coupling}
\label{sec:coupling}
As already outlined, the aerodynamical coupling between the dust and gas phase is related to the grain size and the disc gas mass (see Eq.~\ref{eq:stokes}), i.e.
\begin{equation}
\mathrm{St}_0 \propto \frac{\pi}{2} \frac{a \rho_{\mathrm{d}}}{\Sigma_{\rm gas}} \propto \frac{a}{M_{\rm gas}} .
\label{eq:stokes_midplane}
\end{equation}
where $M_{\rm gas}$ is the gas disc mass.
We have run simulations with initial values of the midplane Stokes number (${\rm St}_0$) in the range $\sim 4\cdot[10^{-3},10^{-2},10^{-1},20]$ at the inner dust radius $R_{\rm in,d}=40$ au, assuming an intrinsic grain density $\rho_{\mathrm{d}}=3\,{\rm g\,cm}^{-3}$. 
In the left panel of Fig.~\ref{fig:stokes} gas surface densities relative to the four choices of Stokes number are displayed, as a function of the disc radius. We note that the Stokes number range used in our models was chosen in order to span a wide range of dust-gas coupling. It is clear that the lower Stokes number of this model is unphysical because it corresponds to a very high disc mass that will lead to the development of gravitational instabilities. 
However, the same study is still relevant for smaller grains and lower intrinsic grain density, which would provide smaller and more physical gas masses for the same degree of coupling. In the right panel of Fig.~\ref{fig:stokes} we show the initial value of the midplane ($z=0$) Stokes number as a function of radius. Each line represents a different gas mass. The solid, dashed and dotted lines (${\rm St}_0\sim 4\cdot[10^{-3},10^{-2},10^{-1}]$) refer to simulations performed with the one-fluid method, while the dot-dashed line ($\rm{St}_0\sim 80$) refers to a two-fluid simulation.

Note that, once the dust is decoupled from the gas (St$\gg 1$), the dust dynamics depends very little on St, so our results for the largest value of St are in general applicable also for moderate St.

\subsubsection{Properties of the embedded planets}
\label{sect:planetprop}
In each disc model we embed two planets at radial distances from the central star of $R_{\rm IP} =35$ au and $R_{\rm OP}=145$ au, respectively. We model the planets and the central star as sink particles. The sinks are free to migrate and are able to accrete gas and dust \citep{bate95}. 
Depending on the planetary mass and local disc structure, a planet can excite multiple spiral arms \citep{miranda18a}.
Generally, planets with masses larger than a threshold, i.e.
\begin{equation}
M_{\mathrm{p}}\gtrsim M_{\rm th} \equiv \left(\frac{H_{\rm g}}{R}\right)_{\mathrm{p}}^3 M_\star  ,
\label{eq:thermalMass}
\end{equation}
are expected to excite two spiral arms interior to their orbit, while the tidal interaction with planets with $M_p\ll M_{\rm th}$ give rise to a spiral structure with only one arm. 
In both cases, the pitch angle of the spiral arms are expected to be proportional to the local aspect ratio $\left(H/R\right)_{\mathrm{p}}$ \citep{rafikov02,zhu15a,miranda18a}.
The mass $M_{\rm th}$ is defined as the planet mass at which the Hill radius of the planet (Eq.~\ref{eq:hill} below) is equal to the thickness of the disc $H(R_{\mathrm{p}})$ at the planet position $R_{\mathrm{p}}$.
The planet masses in our disc models are chosen to be equal to $M_{\rm IP}=3\, M_{\mathrm{j}}$ and $M_{\rm OP}=5\, M_{\mathrm{j}}$, where $M_{\mathrm{j}}$ indicates the Jupiter mass. With this parameter choice, we get an inner cavity and a two armed spiral feature. These masses correspond to $\simeq[23,7,3]\,M_{\rm th}$  and $\simeq[20,6,2.5] M_{\rm th}$ respectively, for $(H/R_{\rm in})_{\rm g}=[0.04,0.06,0.08]$. 
The accretion radius of each planet is chosen to be half the minimum between the Hill radius, 
\begin{equation}
R_{\mathrm{H}} = \left(\frac{1}{3} \frac{M_{\mathrm{p}}}{M_{\star}}\right)^{1/3} R_{\mathrm{p}}  ,
\label{eq:hill}
\end{equation}
and the disc height at the planet position $H(R_{\mathrm{p}})$,
\begin{equation}
R_{\rm acc,p}= 0.5 \min(R_{\rm H},H_{ \rm g}(R_{\mathrm{p}}))  . 
\end{equation}
In this way, we are able to accurately reproduce the spiral arms excited by the planets embedded in the disc, with reasonable computational efforts. Also, since different values of $(H/R_{\rm in})_{\rm g}$ correspond to different values of gas density at the disc midplane, this will result in a slightly different planetary migration and accretion \citep[e.g.][]{dangelo08a,baruteau14a}.

\subsection{Radiative transfer and synthetic observations}
\label{sec:alma}
We compute synthetic observations of our disc models by performing 3D radiative transfer simulations, by means of the \textsc{RADMC-3D} code \citep{dullemond12}, starting from the results of the hydrodynamical simulations. Our goal is to compute the synthetic ALMA Band 6 (1.3 mm) and SPHERE Band H (1.65 $\mathrm{\mu}$m) observations. 
\begin{figure*}
	\includegraphics[scale=0.53,trim={0 8.2cm 0 0},clip]{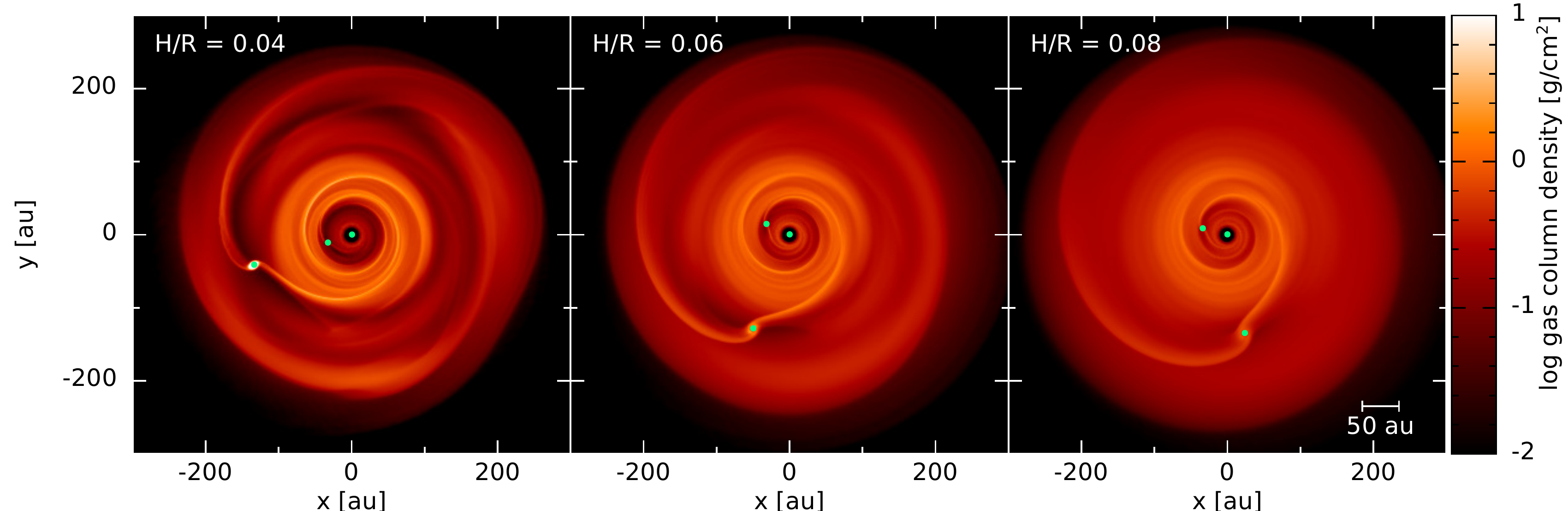}
 	\caption{Gas surface density maps of our set of hydrodynamic simulations. From left to right are shown the results related to disc models with different  aspect ratio $(H/R_{\rm in})_{\rm g}= [0.04,0.06,0.08]$ at the gas inner disc radius $R_{\rm in}=25$ au. 
    }
    \label{fig:density_gas_204}
\end{figure*}
\begin{figure*}
	\includegraphics[scale=0.75,trim={0 0 1.9cm 0},clip]{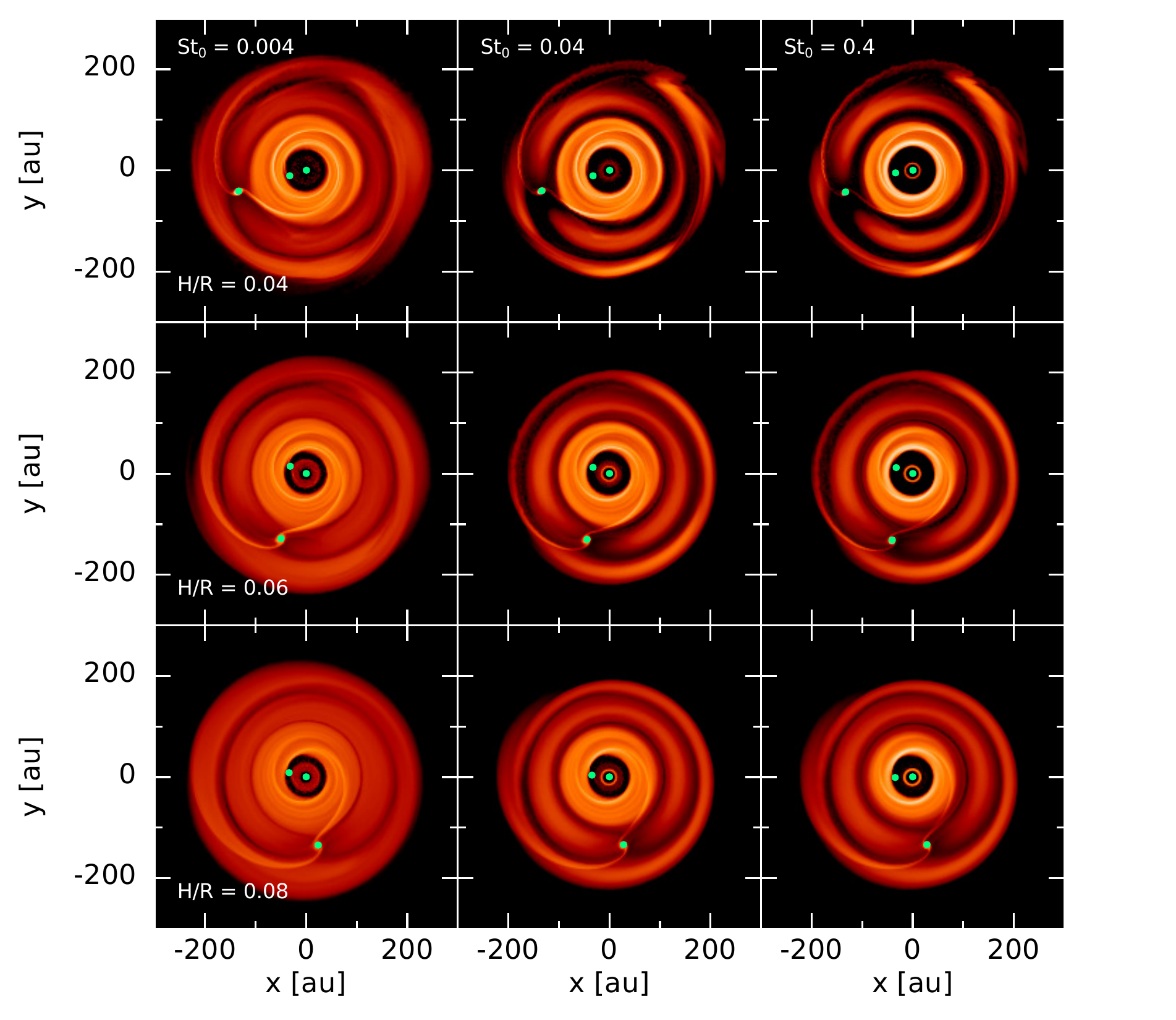}%
    \includegraphics[scale=0.75 ,trim={2.5cm -0.0085cm 0 0},clip]{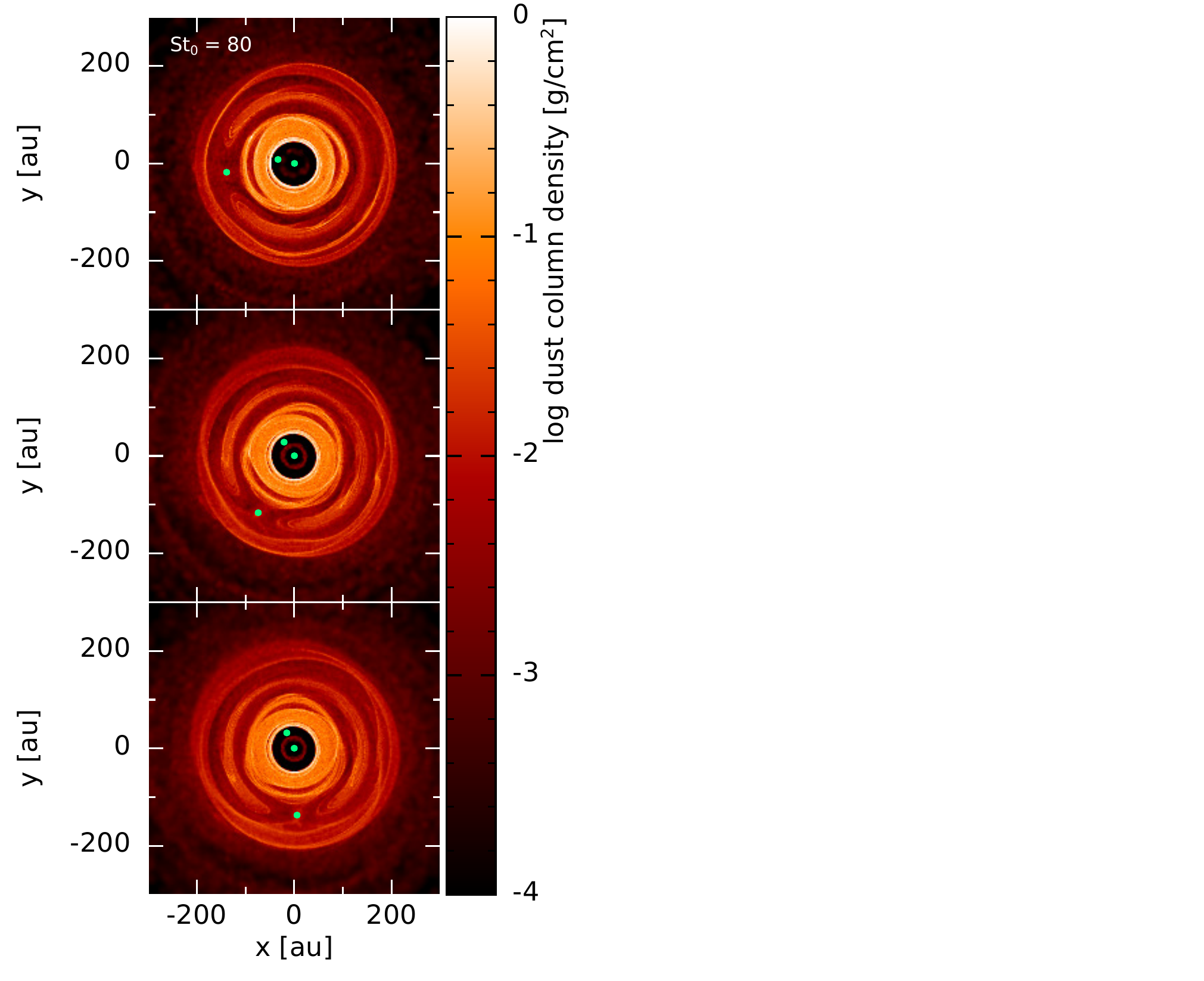} %
 	\caption{Dust surface density maps of our set of hydrodynamic simulations with (from top to bottom) different disc aspect ratio $(H/R_{\rm in})_{\rm g}= [0.04,0.06,0.08]$ at the gas inner disc radius $R_{\rm in,g}=25$ au, and (from left to right) different initial midplane Stokes number $\mathrm{St}_{0}\simeq4\cdot[10^{-3},10^{-2},10^{-1},20]$ at the dust inner disc radius $R_{\rm in,d}=40$ au. 
	}
    \label{fig:density_dust_204}
\end{figure*}

The main inputs for the radiative transfer modelling are the dust density structure of large and small grains, a model for the dust opacities and the source of luminosity. 
We compute the dust opacities using the \textsc{diana} OpacityTool code\footnote{\url{https://dianaproject.wp.st-andrews.ac.uk}} developed by \citet{woitke16}, adopting the dust model from \citet{min16a}.
We choose two ranges of grain sizes in order to sample both the small ($0.1 \lesssim a \lesssim 10 \mu m$) and the large ($0.1 {\rm mm} \lesssim a \lesssim 1 {\rm cm}$) dust population and to study scattered light and continuum images. 
While the large dust ($\sim 1$ mm) is directly available from the simulation, we assume that the small ($\sim 1\,\mu$m) dust density is simply proportional to the gas density (since it is expected to be tightly coupled to the gas for any gas mass), $\Sigma_{\rm sd}(R)= f\Sigma_{\rm gas}(R)$. This factor $f$ is chosen so as to have a total small dust mass of  $M_{\rm sd}\simeq 1.6 \cdot 10 ^{-5}M_\odot$, which corresponds to a ISM-like grain size distribution ($dn/da\propto a^{-3.5}$), if the large dust mass is $M_{\rm ld}\simeq5\cdot 10^{-4} M_\odot$.

The expected emission maps at 1.65 $\mathrm{\mu}$m and 1.3 mm are computed via ray-tracing using $10^9$ photon packages, assuming that the disc is face-on. The source of radiation is assumed to be the central star, located at the centre of the coordinate system, with $M_{\star}= 1.7 M_{\odot}$, $T_{\rm{eff}}=6810$ K and $R_{\star}= 1.4 R_{\odot}$.  We assume the disc to be located in Ophiuchus star-forming region ($d\sim$156~pc). In order to simulate the effect of a coronograph in scattered light observations, we mask a circular region of diameter $\sim 0.15$'' around the central star. This corresponds to the coronograph diameter used in the most common service NIR configuration for the IRDIS instrument\footnote{Table 45 in \url{https://www.eso.org/sci/facilities/paranal/instruments/sphere/doc/VLT-MAN-SPH-14690-0430_v96.pdf}}. 

The full-resolution images at 1.3 mm directly produced by RADMC-3D are used as input sky models to simulate realistic ALMA observations using the Common Astronomy Software Application (CASA) ALMA simulator (version 4.5.3, \citealt{casa}). We take into account the thermal noise from the receivers and the atmosphere \citep{pardo02a} and assume a perfect calibration of the visibility measurements. 
We assumed Cycle 6 ALMA capabilities adopting an antenna configuration alma.cycle6.6 that provides a beam of 0.1 $\times$ 0.08 arcsec ($\sim 16 \times 13$ au at 156~pc), adopting a transit duration of 300 minutes. However, in order to remove possible external asymmetries and to be able to study only non axi-symmetric structures due to the dynamic of the system, we manually changed the beam to a circular one with a dimension of 0.09 $\times$ 0.09 arcsec.
The SPHERE images are computed by convolving the full-resolution scattered light image in H-band produced by RADMC-3D with a circular Gaussian point spread function with a full width half maximum (FWHM) of 0.037'', taken as a good approximation to the angular resolution achieved by SPHERE.

\section{Results}
\label{sec:results}

\begin{figure*}
\includegraphics[scale=0.215]{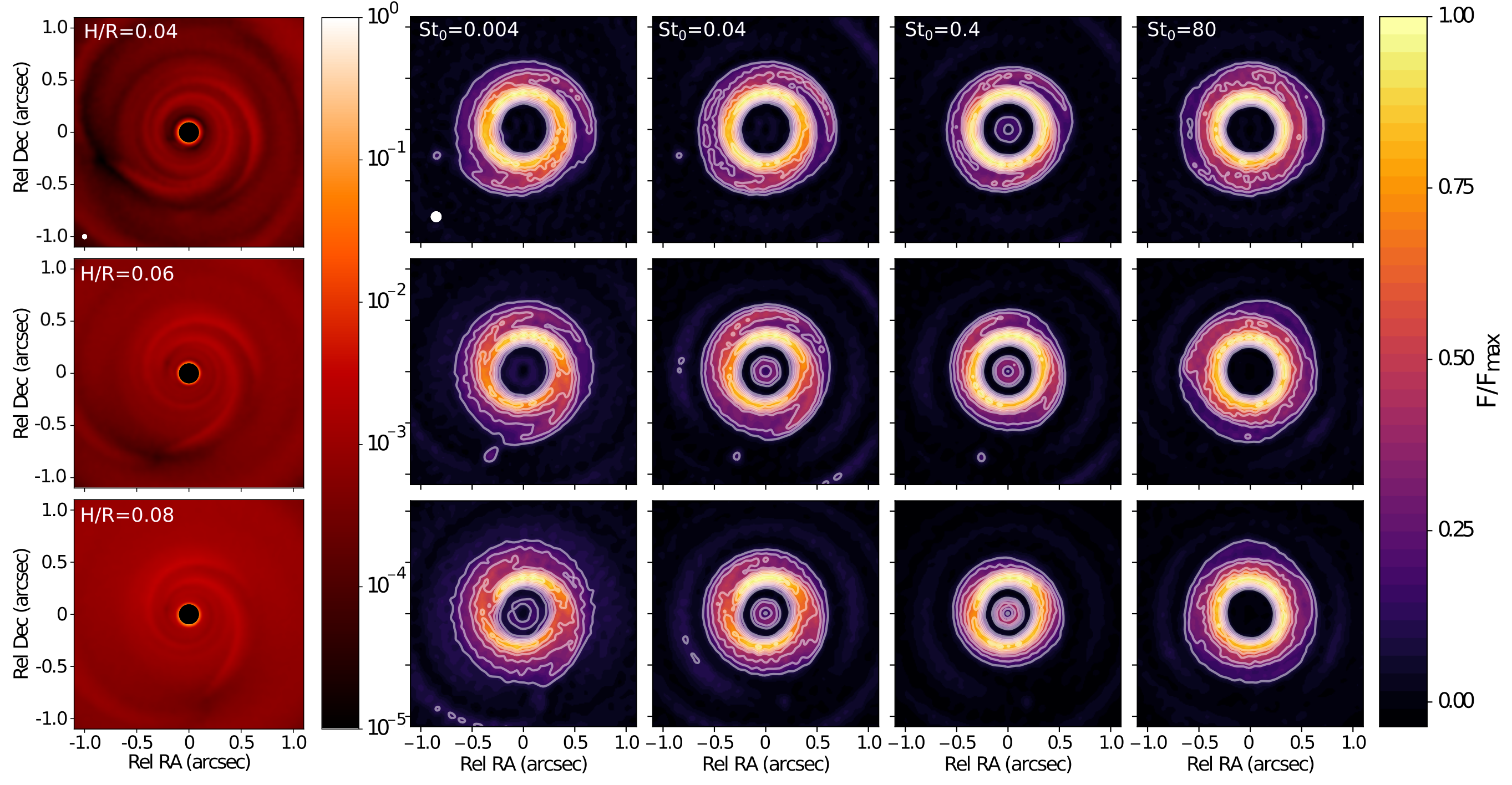}

	\caption{Synthetic SPHERE and ALMA observations of disc models with different disc aspect ratio $(H/R_{\rm in})_{\rm g}=[0.04,0.06,0.08]$, respectively from top to bottom. First column: SPHERE synthetic observations obtained by rescaling the gas surface density into micron-sized dust grains. From the second to the fifth columns are shown the ALMA synthetic images for different Stokes number $\mathrm{St}_0 = 4\cdot [10^{-3},10^{-2},10^{-1},20]$.
	 In the ALMA images contours are $5,10,15,20,25,30,35,40,45,50,55,60$ times the rms noise ($2.5\cdot 10^{-2}\,$mJ) at 230 GHz. Both the ALMA and SPHERE images have been renormalized to the maximum value of the flux. The SPHERE images have a logarithmic flux scale.}
    \label{fig:mock_images}
\end{figure*}

\subsection{Disc morphologies as a function of the Stokes number}
\label{sect:discmorph}
Figure~\ref{fig:density_gas_204} shows the gas surface density maps of our disc models with the three aspect ratio (from left to right) $(H/R_{\rm in})_{\rm g}=[0.04,0.06,0.08]$ after $\sim 10$ orbits of the outer planet (at the initial planet location $R_{\mathrm{OP}}=145$ au) and 80 orbits of the inner planet (initial location $R_{\mathrm{IP}}=35$ au). Due to the different locations and masses of the embedded planets, the resulting spiral structures induced by planet-disc interactions is dominated by the outer planet, while the inner one is responsible for carving the cavity.
As already mentioned in Sect.~\ref{sect:planetprop}, the spiral features interior to the orbit of the outer planet are characterized by two spiral arms with a pitch angles that increases with increasing $(H/R_{\rm in})_{\rm g}$. At the same time, the arms become less sharp, since the intensity of the spiral structures depends on the planet mass and on the disc aspect ratio (eq. 16 in \citealt{miranda18a}). The density colour scale is in arbitrary units and the gas densities for the various choices of disc mass can be obtained by a suitable rescaling.

Fig.~\ref{fig:density_dust_204} shows the dust surface density maps of our models. Here the different rows represent different disc aspect ratio $(H/R_{\rm in})_{\rm g}=0.04,0.06,0.08$, while each column represents a specific Stokes number at the disc midplane ($\rm{St}_0=4\cdot[10^{-3},10^{-2},10^{-1},20]$ at the reference radius $R_{\rm in,d}$ for the initial conditions).
All the snapshot of Fig.~\ref{fig:density_dust_204} are taken at the same time, so the different position of the planets are due to a slightly different migration rate for the various disc aspect ratios. 
If we look at the last column of Fig.~\ref{fig:density_dust_204}, $\rm{St}_0=80$, some elliptical asymmetries are visible. These are caused by the fact that if the dust is decoupled from the gas, eccentricity pumping at the mean motion resonances (i.e. 1:2, 2:1, 2:3 and 3:2, which are respectively 0.6, 1.6, 0.76 and 1.3 times the planet radius $R_p$) can arise \citep{zhu14a}. The inner resonances (1:2 and 2:3) in our case are at $\simeq 87$ au and $\simeq 110$ au, that correspond with the elliptical shape visible in the dust with $\rm{St}_0=80$.

\begin{figure*}
\includegraphics[scale=0.63]{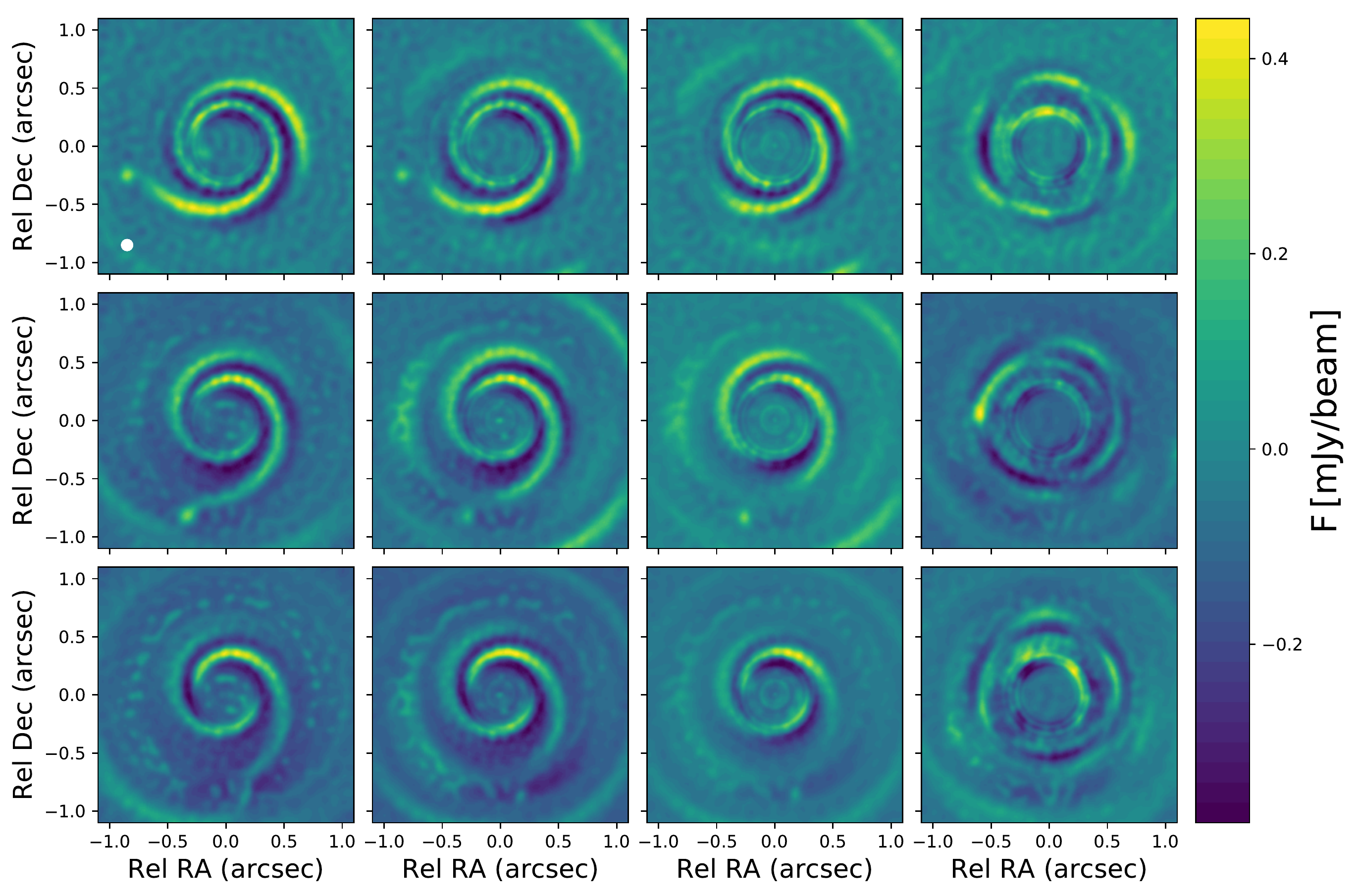}
	\caption{Residuals of the synthetic ALMA observations shown in Fig.~\ref{fig:mock_images}, for $(H/R_{\rm in})_{\rm g}=0.04,0.06,0.08$ (top to bottom) and for different Stokes number $\mathrm{St}_0 = 4\cdot [10^{-3},10^{-2},10^{-1},20]$ (left to right). To obtain these residuals we divide the disc in 22 annuli from 0.02 to 1.0 arcsec. Then, we subtract from the intensity of each annulus the respective mean intensity.}
    \label{fig:residuals}
\end{figure*}

\subsection{Simulated images}
\label{sec:obs}
When the dust and the gas are coupled (i.e. for high gas mass) we expect to detect spiral structures, both in the dust continuum and in the scattered light emission, while when they are decoupled (i.e. lower gas mass) the disc will be characterized by a more axi-symmetric structure at mm-wavelengths (e.g. rings or horseshoes). Fig.~\ref{fig:mock_images} shows the synthetic images obtained as described in Section ~\ref{sec:alma} for the SPHERE and ALMA observations. In the first column there are the scattered light SPHERE images (micron-sized grains).
The dust continuum ALMA images (tracers of millimetric grains distribution) are presented for different Stokes parameters ($\mathrm{St}_0=4\cdot[10^{-3},10^{-2},10^{-1},20]$ at the midplane in the initial condition) from the second to the fifth columns. Different disc aspect ratio $(H/R_{\rm in})_{\rm g}=0.04,0.06,0.08$ are shown from the top to the bottom row.

As expected all SPHERE images show a spiral feature, since we obtained them by rescaling the gas by a factor that takes into account the amount of micron-sized dust in the disc. By doing in this way, the dynamics of the micron-sized grains are assumed to follow the gas, producing the characteristic spirals we see. 
The ALMA images with $\mathrm{St}_0=4\cdot 10^{-3}$ (for all the disc aspect ratio), also display two spiral arms as visible also in the SPHERE image. Indeed, in this case the dust is strongly coupled to the gas. However, increasing the Stokes number (i.e. decreasing the gas disc mass) the dust and gas become progressively less coupled: the non axi-symmetric sub-structures become gradually more symmetric and circular. As an example, if we consider the ALMA image with $\mathrm{St}_0=80$ and we compare it with the $\mathrm{St}_0=4\cdot 10^{-3}$ image, a ring-like structure instead of the two-armed spiral is observed. 
From St$_0=4\cdot10^{-3}$ to St$_0=0.4$ the disc also becomes more compact since radial dust drift becomes more effective \citep{weiden77}. This effect has been also discussed by \cite{powell19}. At St$_0=80$ the disc retains its original radius due to the strong decoupling between dust and gas.

If we look at different disc aspect ratios, while the spiral structures are well defined in a disc with $(H/R_{\rm in})_{\rm g}=0.04$, they become less evident in a disc with $(H/R_{\rm in})_{\rm g}=0.08$ also for low Stokes number. In fact, when the disc is thinner, since the interaction between the planet and the disc is stronger, the gap carved by the planet is deeper and the resulting spiral is brighter with respect to the background (see Fig. 1 in \citealt{fung15}). We highlight that increasing the aspect ratio the spiral arms become less wound up. This is visible especially for the case with initial $\mathrm{St}_0=0.4$.

\section{Discussion}
\label{sec:discussion}
\subsection{Spiral or ring? An analysis of substructure (a)symmetries}
\label{sect:spirals_disc}
In Fig.~\ref{fig:mock_images} we have shown the ALMA synthetic images for different gas masses, corresponding to different Stokes number. However, since the dynamic range of the image is large, in order to enhance the fainter non axi-symmetric features, we compute the residuals of each image. First, we divide the disc in 22 annuli from 0.02 to 1.0 arcsec. The chosen radial width $\Delta R=0.045''$ corresponds to the beam radius, and has been chosen as a fair compromise between the resolution of our images and the need of thin annuli. Then, in each annulus $j$, we subtract from the pixel intensity $ I_{\rm pix,j}$ the flux mean value $\overline{I_{\rm ann,j}}$, as 
\begin{equation}
    \overline{I_{\rm pix,j}} =  I_{\rm pix,j} - \overline{I_{\rm ann,j}}\,.
\end{equation}
The result is shown in Fig.~\ref{fig:residuals}, where the Stokes number increases from left to right and the disc aspect ratio increases from the top to the bottom. In this figure the more axi-symmetric component has been removed, and the spirals, if present, are clearly visible. As already evinced from Fig.~\ref{fig:mock_images}, the disc with low Stokes number and smaller aspect ratio $H/R$ are characterized by spirals. Increasing the Stokes number the spirals become more tightly wound and also smaller, until they disappear in the disc with the higher Stokes number. However, for St$_0$=80 the residual images appear to be less symmetric than expected. This could be due to the fact that in the starting image were already present some asymmetries (eccentricity pumping, e.g. \citealt{zhu14a}), and by computing the residuals we are enhancing them. 

\begin{figure}
\begin{center}
	\includegraphics[scale=0.4]{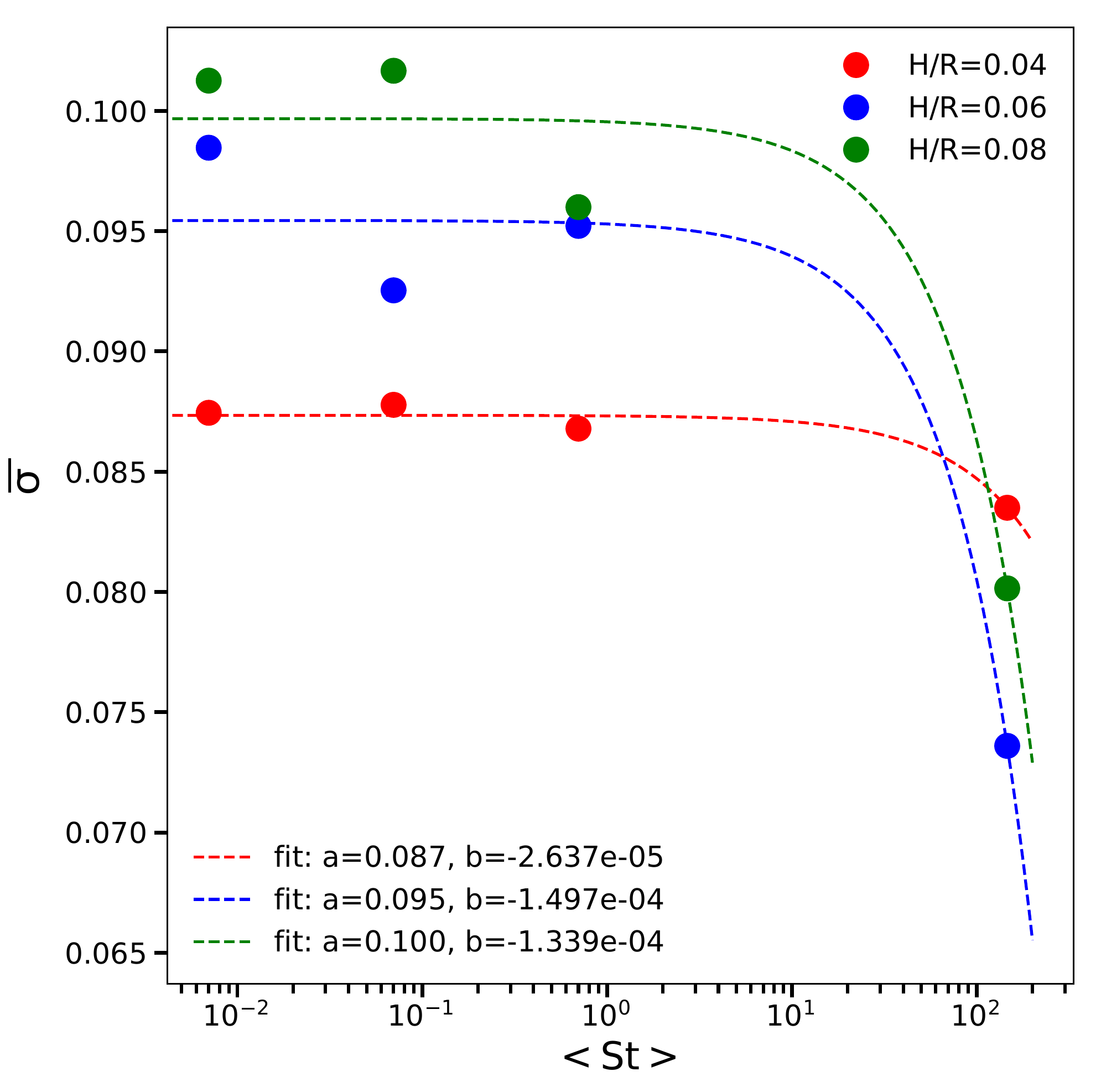}
 	\caption{Weighted average of the standard deviation for the whole disc, as defined in Eq.~\ref{eq:stddev}, as a function of the Stokes number (in log scale) for different disc aspect ratio $(H/R_{\rm in})_{\rm g} = [0.04,0.06,0.08]$ (red, blue and green points-lines respectively). The weight is the square mean intensity computed in the 22 annuli between 0.02 and 1.0 arcsec. The dashed lines represent the fit of the data with first order polynomials.
     }
    \label{fig:dev_std_st}
\end{center}
\end{figure}

We now apply a more quantitative method in order to relate the morphologies (i.e. asymmetries or symmetries) of the residuals images (Fig.~\ref{fig:residuals}) to the gas disc mass. 
We compute for each image of Fig.~\ref{fig:mock_images} the standard deviation of the intensity for each annulus.
The flux in each pixels, $I_{i}$, has been normalized to the maximum value for each images, $I_{i}=I_{i} / {\rm max}{( I) }$. 
Then, we compute the weighted average of the standard deviation for the whole disc, where the weight is the square mean intensity of each annulus,
\begin{equation}
    \overline{\sigma} = \frac{\sum_{j=1}^{N_{\rm ann}}{\sigma_{j} \overline{I_j}^{2}}}{\sum_{j=1}^{N_{\rm ann}}{\overline{I_j}^{2}}}\, .
    \label{eq:stddev}
\end{equation}
In this equation, $\sigma_{j}$ is the standard deviation and $\overline{I_j}$ is the mean flux, in each annulus. In Figure~\ref{fig:dev_std_st} we show the weighted average standard deviation as a function of the Stokes number, where the points represent the value computed from the residuals of the simulated images (red for $(H/R_{\rm in})_{\rm g}=0.04$, blue for $(H/R_{\rm in})_{\rm g}=0.06$ and green for $(H/R_{\rm in})_{\rm g}=0.08$), while the dashed lines are the first order polynomial fit of the data points. 
Combining this result with Fig.~\ref{fig:mock_images}, it is interesting to note that the general trend of structures becoming more symmetric with increasing Stokes number is visible also through the standard deviation of the intensity. In addition, by increasing the disc aspect ratio, the standard deviation of the flux intensity increases slightly. 

\begin{table*}
\caption{Sources name, age of the disc, dust disc mass and disc morphologies for the protoplanetary discs discussed in Sec~\ref{sect:application}.}
\label{table:Mstar_HR_Mdisc} 

\begin{threeparttable}
\scalebox{0.83}{
\begin{tabular}{lcccc}
 \hline
Source name & $t\,[\rm{Myr}]$ &  $M_{\rm d,disc}\,[M_{\odot}]$ & Ref. (Age, $M_{\rm d,disc}$) & Morphologies \\ 
(a)&(b)&(c)&(d)&(e) \\
\hline
\smallskip
HD135344B & 8 & $ [1-4]\cdot 10^{-4}$ & [\cite{grady09} - \cite{cazzoletti18}] & A$=$horseshoes+ring; S$=$spirals$\,^{1}$\\ 
TWHya & 10 & $ [2-6] \cdot 10^{-4}$   & [\cite{Weinberger13} - \cite{calvet02,thi10}]  & A,S$=$ rings$\,^{2}$ \\
MWC758 & $3.5^{+2}_{-2}$ &   $ 2\cdot10^{-4}$    & [\cite{meeus12} - \cite{boehler18}] & A$=$ring+spiral/clumps; S$=$spirals$\,^{3}$\\
HD97048 & 3 & $ 7 \cdot 10^{-4}$ & [\cite{Lagage06} - \cite{vanderplas17}] & A,S$=$ rings$\,^{4}$\\
AB Aurigae & 4. & $ 7.5\cdot 10^{-5} $ & [\cite{dewarf03} - \cite{lin06}]& A$=$rings,spirals;S$=$spirals$\,^{5}$\\
PDS70 &  5.4 &  $ 3\cdot10^{-5}  $ & [\cite{muller18} - \cite{keppler18}]& A$=$rings;S$=$rings$\,^{6}$\\
IMLup &  0.5 &   $ 3\cdot10^{-6}  $ & [\cite{alcala17} - \cite{avenhaus18}]& A$=$spirals;S$=$rings$\,^{7}$\\
HD169142 & $ 6^{+6}_{-3}$ &   $ 10^{-4}  $ & [\cite{grady07} - \cite{fedele17}]& A$=$rings;S$=$rings$\,^{8}$\\
HD 143006 &  $ 11.9^{+3.7}_{-5.8}$ &   $ 0.39 \cdot 10^{-4}$ & [\cite{garufi18a} - \cite{natta04}]& A$=$rings;S$=$rings, clump$\,^{9}$\\
\hline
\end{tabular}
}

(a) Source name; (b) Disc age; (c) dust disc mass; (d) references for the disc age and for the dust disc mass; (e) disc morphologies observed in the continuum, with ALMA (A) and in scattered light, with SPHERE-VLT/Subaru-HiCIAO (S). \\ Continuum/scattered light discs morphologies references: \footnotesize{$^1$ \cite{cazzoletti18,maire17}, $^2$ \cite{vanboekel17,andrews16}, $^3$ \cite{boehler18,benisty15}, $^4$ \cite{vanderplas17,ginski16}, $^{5}$ \cite{Hashimoto11}, $^{6}$ \cite{longz18,keppler18}, $^{7}$ \cite{huang18,avenhaus18},$^{8}$ \cite{fedele17,pohl17},$^{9}$ \cite{perez18,benisty18} }\\
\end{threeparttable}
\end{table*}

\subsection{An application to observed morphologies in protoplanetary discs}
\label{sect:application}
The main result of our work is that disc substructures due to planet-disc interaction should be characterized by rings or ring-like structures, rather than spirals, if the Stokes number of millimetre-emitting grains ($a_{\rm max}=\lambda/2\pi$, e.g. \citealt{kataoka15}) is larger than $\sim 0.4$. Provided that the properties of the dust (fluffiness, porosity, intrinsic grain density and size) are known, we can infer the gas surface density, which corresponds to $\Sigma_{\rm gas}\lesssim 0.4 \mbox{g/cm}^2$ (using Eq.~\ref{eq:stokes_midplane} and assuming $\rho_{d}=3\rm{gcm}^{-3}$), and hence to a disc mass of the order of $\approx 1.4 \cdot 10^{-3}M_{\odot}$ for structures located at $\approx 100$ au or to $\approx 5 \cdot 10^{-3}M_{\odot}$ for structures located at 200 au (indeed note, for example, that \citealt{dipierro15} have used a mass of $2\cdot10^{-4}M_{\odot}$ within 120 au to reproduce the system of rings in HL Tau). What implications does this have, when one takes into account the observed disc structures? First of all, we note that the majority of substructures observed with ALMA at high resolution are in the form of rings \citep{huang18,long18}, which would then point to a prevalence of low mass discs. Note, however, that dust rings may be produced by mechanisms other than planet-disc interaction, so that the mere detection of rings in the mm continuum does not in itself imply the presence of a planet. A stronger indication for planet-disc interaction is present in those cases where, in addition to structure in the mm continuum, one also observe a substructure, such as a spiral, in scattered light images. We thus list in Table 2 all the known discs that have shown substructures both in scattered light and in mm continuum images, indicating for each of them what kind of structure is observed. From this small sample of discs, it is clear that in order to build a wider picture of disc evolution and to unravel the gas disc mass problem, the observation of a wider zoo of protoplanetary discs at different wavelengths is crucial \citep{haworth16a}.
However, we can extract interesting clues from the discussion of some of these systems.

In some cases, such as HD135344B and MWC758, we indeed observe a spiral structure in scattered light and rings or circular structures in the mm continuum, indicating that the disc mass in these system should be relatively low. In particular, for the case of HD135344B, which was the initial motivation of our study, we confirm that the observed spiral structure in scattered light can be ascribed to the presence of two massive planets, with a mass of a few Jupiter masses. ALMA images of HD135344B show a circular structure, but rather than full rings, a horseshoe is observed \citep{cazzoletti18}, which might indicate the presence of a vortex \citep{vandermarel16}, that we cannot reproduce in our images because the viscosity that we assume is too large for vortex production. In MWC758 we do see hints of spiral structures: a southern arm in the continuum emission (optically thin, traces disc midplane) and two of them in the $^{13}$CO J=3-2 peak emission lines (optically thick, traces variation of the disc temperature), almost in the same location of the ones observed in the scattered light (see \citealt{boehler18,dong18a}). The small offset in the radial location of the southern spiral observed in the continuum is probably due to the vertical propagation of the spiral (i.e. they bend over toward the star), or to a nonzero inclination so that the ``NIR/surface spiral" and ``ALMA/midplane spiral" are projected to different locations on the plane of the sky. The origin of the  spiral observed in the continuum could be found in the trapping of millimeter-sized dust particles. Two clumps in correspondence of the two NIR spirals are also observed, and they are explained as possible Rossby wave instability \citep{boehler18}. Also \citet{dong18a} hypothesized that their radial widths is related to emitting particles having Stokes numbers smaller than unity. This could be due to a locally high gas surface density.

TW Hya, HD143006 and HD169142 are three old ($t\geq 6-12$ Myr, \citealt{Weinberger13,grady07,garufi18a}) protoplanetary discs with symmetric structures imaged both in the ALMA and in the SPHERE data \citep{vanboekel17,andrews16,fedele17,pohl17}. They all are evolved systems, where the gas has almost all been dissipated leaving a disc with a low gas-to-dust ratio \citep{williams11}. This low surface gas density implies a high Stokes number not only for the millimetre grains, but also for the micron-sized dust, producing the observed structures in the SPHERE images. For this to happen, the gas disc mass should be smaller than $10^{-4}M_{\odot}$. The case of TW Hya is particularly controversial. While its age would be in line with a very small gas mass, HD line observations by \citet{bergin13} would instead point to a much higher gas mass (i.e. $M_{\rm gas}>0.06\,M_{\odot}$). Note that recently \citet{powell19} have performed an analysis similar in spirit to our own, in that they try and weigh the gas mass by estimating the different degrees of dust-gas coupling comparing multi-wavelength images. However, rather than using the effect of dust coupling on sub-structures, as we do, they observe its effect on the dust disc size, whereby dust with a larger St is expected to drift at a faster rate and produce a smaller disc (an effect that we also see in our simulations). By using this method, they generally conclude that the proto-planetary discs that they have analysed are relatively massive (close to gravitational instability). In particular, for TW Hya, they estimate a mass of 0.11 $M_{\odot}$, in line with the HD measurements. However, this beg the question of how, with such a high disc mass, a regular system of rings is observed in the ALMA continuum. \citet{mentiplay19} have recently modeled TW Hya using \textsc{phantom} and a gas mass $\lesssim 10^{-2} M_{\odot}$ and indeed find that the scattered light images of their best model do show a spiral structure. To produce rings also in scattered light would require a very low disc mass $\lesssim 10^{-5} M_{\odot}$, well below the minimum of $3 \cdot 10^{-4} M_{\odot}$ inferred from CS measurements by \citealt{teague18}.

Another interesting case is IM Lup, for which a tightly wound spiral structure is observed with ALMA \citep{huang18}, while in scattered light the structure could be described both as multiple rings (Avenhaus et al. 2018) or as another tightly wound spiral. Despite a very low amount of dust (Avenhaus et al. 2018), this is thought to be one of the largest disc with a radial extent of $\approx 950$ au, measured from CO data, and a gas mass of $0.17 M_{\odot}$ \citep{pinte18}\footnote{\cite{ansdell18a} found a gas mass of $10^{-2}\,M_{\odot}$, which is lower than the one given by \cite{pinte18}. This discrepancy highlights the fact that we still have a large uncertainty on the gas mass measurements.}. This would be consistent with a spiral induced by planet disc interaction in a massive disc with small $H/R$ (in order for the spiral to be tight). However, note that in this parameter range, the disc would be also prone to gravitational instability, which may be responsible for the observed spiral. 

\subsection{Caveats}
Our calculations consider only $\simeq$ 10 orbits of the outer planets, due to computational resources. However, it is worth highlighting that the relation we found between the observed substructures and the disc Stokes number is time-dependent: in more evolved systems the gas has already been partially dissipated, involving lower Stokes number and more symmetric structures. Also, the gap carving process is determined by the viscous timescale and for lower Stokes number it requires a large number of orbits. This means that while the transition between spirals and more axisymmetric structures in the continuum is general, the threshold between the two regimes may be different with longer timescale simulations (i.e. $\rm{St}_{0}<0.4$).

From an observational point of view, to be able to distinguish between spiral and ring in scattered light one needs large pitch angle spiral, such the one in HD135344B or MWC758, unlike the one observed in TW Hya.

Moreover, it is important to note that in a single protoplanetary disc we can have regions with high and low Stokes number, resulting in a locally different coupling and so in different substructures. This can be due to both a lower/higher surface density regions (along the radial/vertical direction) and to different grain size populations in the same disc. In the first scenario, this is the result of the disc-planet interaction, with the formation of gaps and/or cavity \citep{dipierro16a,ragusa17}, and of photo-evaporative process \citep{alexander14}, while in the second one this is the result of the process of grain growth and settling \citep{testi14a}. For this reason, in order to obtain a more accurate study of the system, one should consider a single model with a wide range of dust grains sizes \citep{hutchinson18, dipierro18dust}. This is now possible with the new \textsc{multigrain} utility that has been developed in the \textsc{phantom} code. 

\section{Conclusions}
\label{sec:conclusion}
One of the most crucial, unsolved, questions in the field of planet formation is ``how massive are protoplanetary discs?" \citep{bergin18a}. Indeed, the gas mass affects how planets form and what is the origin of the substructures we observe in protoplanetary discs. In this paper we explore if there is a relationship between the observed morphologies (spirals, gaps, rings) and the gas mass. We have performed 3D dust and gas Smoothed Particle Hydrodynamics and radiative transfer simulations of protoplanetary discs with embedded planets, exploring different aerodynamical coupling between millimetre dust grains and gas and different aspect ratios. 

The basic result of our work is that, as a result of planet-disc interaction, spiral structure observed in scattered light images would be observed as a circular, ring-like configuration in ALMA continuum emission for Stokes number close to unity or larger, corresponding to gas surface density $\lesssim 0.4 \mbox{g}/\mbox{cm}^2$. The observed prevalence of ring-like structures in ALMA images would thus imply that discs are generally less massive than $\approx 10^{-3}\,M_{\odot}$. \citet{powell19} use a method similar to ours to estimate the gas disc mass but use the dust radial extent rather than the dust morphology to estimate the Stokes number. By using their method, they obtain an opposite result, in that most of the disc that they have considered would instead result in very massive discs, often close to gravitational stability. Further work is needed to resolve this discrepancy.

\section*{Acknowledgements}
We wish to thank the referee for an insightful report of the manuscript. We thank Nienke van der Marel for fruitful discussions that gave new perspective to this work. We also want to thank Anna Miotello for enlightening discussion about gas measurements from observations.
GL and BV have received funding from the European Union’s Horizon 2020 research and innovation programme under the Marie Skłodowska-Curie grant agreement No 823823 (RISE DUSTBUSTERS project).
GD and ER acknowledge financial support from the European Research Council (ERC) under the European Union's Horizon 2020 research and innovation programme (grant agreement No 681601). CH is a Winton Fellow, and this research has been supported by Winton Philanthropies/The David and Claudia Harding Foundation. DP acknowledges funding from the Australian Research Council via FT130100034 and DP180104235.

For our work we adopted different codes and software. We used \textsc{phantom} \citep{price18phantom} for the hydrodynamic simulations and \textsc{radmc-3d} \citep{dullemond12} for the radiative trasnfer calculations. We utilized \textsc{diana OpacityTool} \citep{woitke16} for the dust opacities coefficients and \textsc{casa} \citep{casa} for the simulations of ALMA observations. We used \textsc{SPLASH} \citep{splash} for rendered images of our simulated hydrodynamic systems, while the remaining figures have been generated using the \textsc{python}-based MATPLOTLIB package  \citep{matplotlib}.


\bibliographystyle{mnras}
\bibliography{biblio} 




\bsp	
\label{lastpage}
\end{document}